\documentclass[twocolumn,twoside,nofootinbib]{revtex4-1}
\usepackage{graphicx}
\usepackage[percent]{overpic}
\usepackage{fancyhdr}
\newcommand{\Bsmumu}{\ensuremath{B_s^0 \to \mu^+\mu^-}}
\newcommand{\Bdmumu}{\ensuremath{B^0 \to \mu^+\mu^-}}
\newcommand{\Bsdmumu}{\ensuremath{B_{(s)}^0 \to \mu^+\mu^-}}

\newcommand{\BpmKpmJpsi}{\ensuremath{B^\pm \to J/\psi K^\pm}}
\newcommand{\BsJpsiPhi}{\ensuremath{B_s^0 \to J/\psi\phi}}

\newcommand{\JpsiMuMu}{\ensuremath{J/\psi \to \mu^+\mu^-}}
\newcommand{\Bhh}{\ensuremath{B \to hh^\prime}}
\newcommand{\Bsdhh}{\ensuremath{B^0_{(s)} \to hh^\prime}}

\newcommand{\Bs} {\ensuremath{B_{s}^0}}
\newcommand{\Bsbar} {\ensuremath{\bar{B}_{s}^0}}
\newcommand{\Bd} {\ensuremath{B^0}}

\newcommand{\Bsd} {\ensuremath{B_{s,d}^0}}

\newcommand{\mumu}{\ensuremath{{\mu^+\mu^-}}}
\newcommand{\JpsiK}{\ensuremath{{J/\psi K^\pm}}}
\newcommand{\Jpsi}{\ensuremath{{J/\psi}}}

\newcommand{\bmumuX}{\ensuremath{b \to \mu^+\mu^- X}}

\newcommand{\Phikk}{\ensuremath{\phi \to K^+K^-}}

\newcommand{\ifb}{\mbox{fb$^{-1}$}}

\newcommand{\BR}{{\ensuremath{\cal B}}}

\newcommand{\DeltaLnLike}{\ensuremath{\Delta\ln\mathcal{L}}}

\newcommand{\mmumu}{\ensuremath{m_{\mumu}}}

\newcommand{\mJpsiK}{\ensuremath{m_{\JpsiK}}}

\newcommand{\pt}{\ensuremath{p_{T}}}

\begin{document}

\title{ATLAS Measurements of CP Violation and Rare Decays in Beauty Mesons}

%

\author{W. Walkowiak\\
  on behalf of the ATLAS Collaboration}
\affiliation{University of Siegen, 57068~Siegen, Germany}

\begin{abstract}
  The ATLAS experiment at the Large Hadron Collider (LHC) 
  has performed accurate measurements of mixing
  and CP violation in the neutral B mesons, and also of rare processes
  happening in electroweak FCNC-suppressed neutral B-mesons decays.
  This contribution focuses on the latest results from ATLAS,
  including measurements of 
  rare processes \Bsmumu\ and \Bdmumu, and
  measurements of CP violation in \BsJpsiPhi.
\end{abstract}

\maketitle
\thispagestyle{fancy}


\section{Introduction}

New physics beyond the Standard Model (SM) may manifest itself in
the branching fractions of very rare $B$ meson decays or CP-violating
parameters in \Bs\ oscillations.
The ATLAS experiment~\cite{ATLASdet}
at the Large Hadron Collider (LHC)~\cite{LHCpaper} at CERN performs
indirect searches for New Physics by measuring
the branching fractions of the rare decays \Bsmumu\ and
\Bdmumu\ 
and the CP-violating phase $\phi_s$ as well as $\Delta\Gamma_s$
in the \BsJpsiPhi\ decay.
In addition, projections for the branching fractions of the rare
decays \Bsdmumu\ and expected sensitivities for the search for
CP violation in the decay channel \BsJpsiPhi\ at the High-Luminosity
LHC (HL-LHC)~\cite{ybHLLHC} are
presented\footnote{Copyright 2019 CERN for the benefit of the ATLAS
  Collaboration CC-BY-4.0 license.}.

\section{Branching fractions of \Bsmumu\ and \Bdmumu}

\begin{figure}[t]
  \centerline{
    \includegraphics[width=80mm]{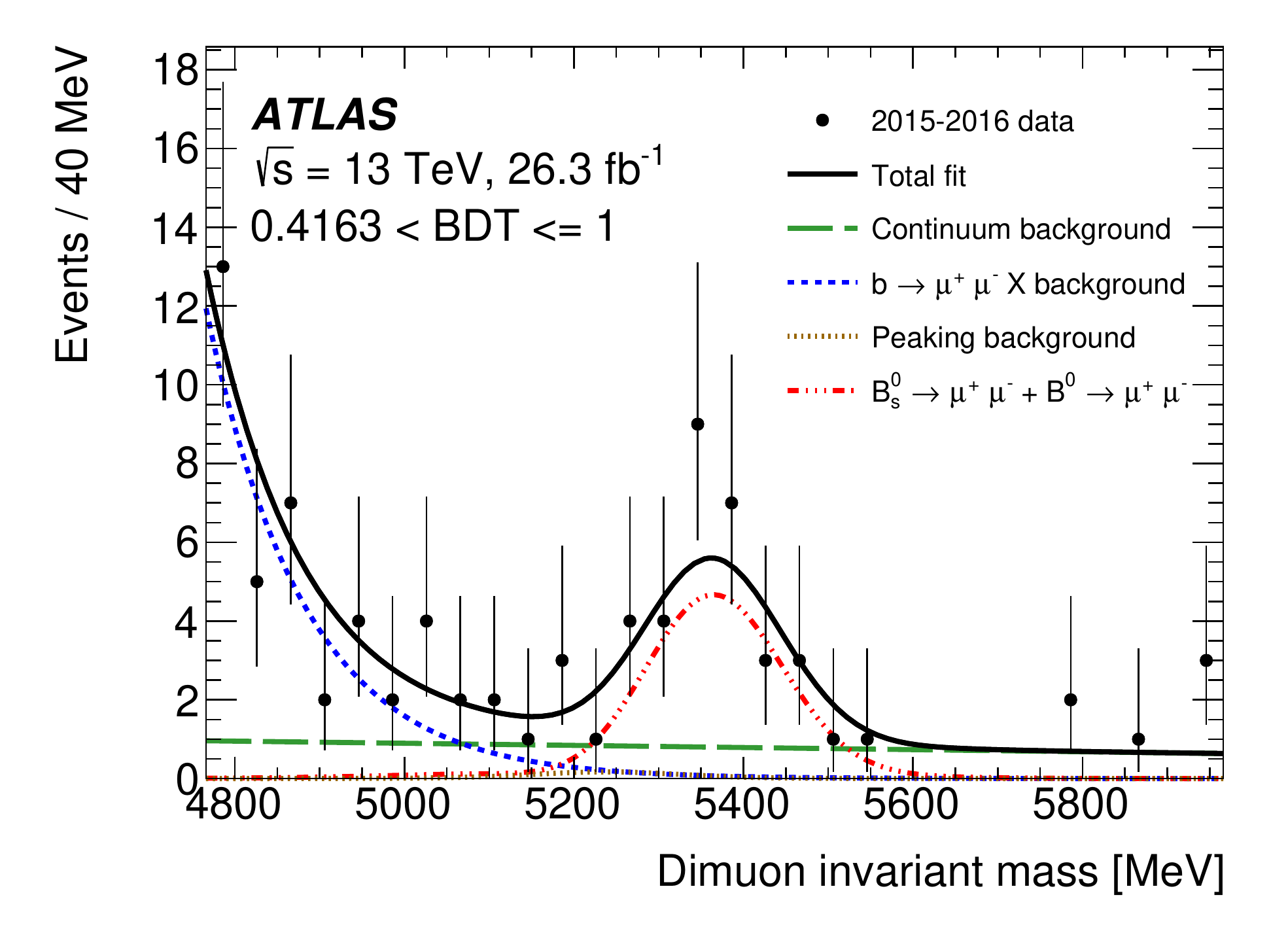}
  }
  \caption{\label{fig:bmumurun2mass4}
    Dimuon invariant mass distribution in the unblinded data, for the
    highest interval of BDT output.
    The result of the maximum-likelihood fit is superimposed.
    The total fit is shown as a
    continuous line, with the dashed lines corresponding to the
    observed signal
    component, the \bmumuX\ background, and the continuum
    background. The signal components are grouped in one single curve,
    including both the \Bsmumu\ and the (negative) \Bdmumu\ 
    component. The curve representing the peaking \Bsdhh\ 
    background lies very close to the horizontal axis~\cite{ATLASBmumu2019}.
  }
\end{figure}

\begin{figure}[t]
  \centerline{
    \includegraphics[width=80mm]{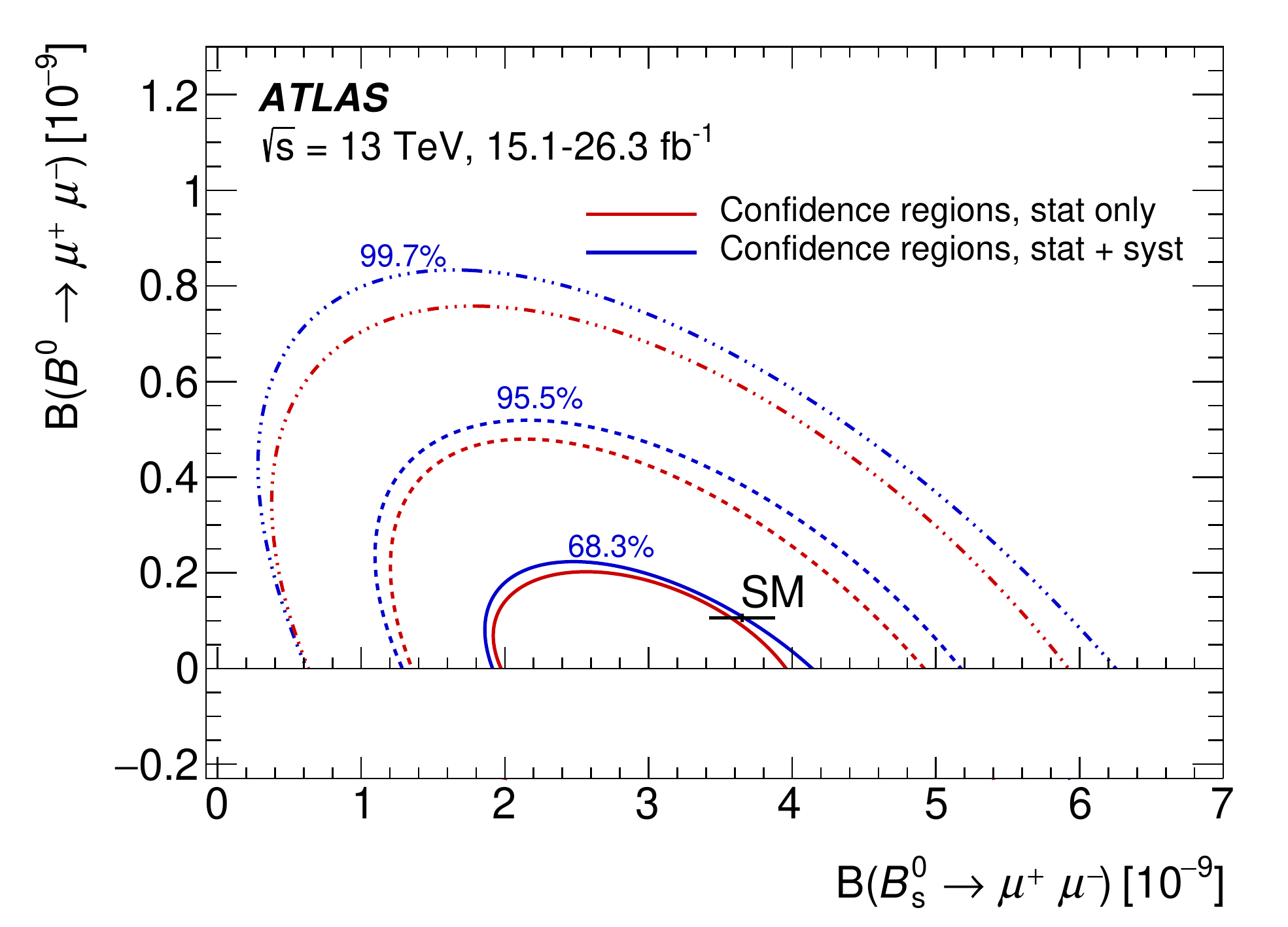}
  }
  \caption{\label{fig:bmumurun2neym}
    Neyman contours in the $\BR(\Bsmumu)$ - $\BR(\Bdmumu)$ 
    plane for 68.3, 95.5 and 99.7 \% coverage. At each
    coverage value, the inner contours are statistical uncertainty
    only, while the outer ones include statistical and systematic
    uncertainties. The construction of these contours makes use of
    both the dimuon (26.3~\ifb) and the reference channel (15.1~\ifb)
    datasets~\cite{ATLASBmumu2019}.
  }
\end{figure}

The rare decays \Bsmumu\ and \Bdmumu,
which are sensitive to New
Physics in the decays via loop diagrams,
are highly suppressed in the
Standard Model (SM) with predicted branching
fractions~\cite{bobeth2014}
of $(3.65\pm 0.23)\times 10^{-9}$ and $(1.06\pm 0.09)\times 10^{-10}$,
respectively. 
The ATLAS Run~1 result~\cite{atlasbs2016} is compatible with
the SM at $\sim 2\sigma$ level, and the \BR(\Bsdmumu) values are lower
than the CMS-LHCb combined result~\cite{CMSLHCbcomb2015}.
A recent LHCb measurement~\cite{LHCbBmumu2017} including a part of Run~2 data
sets an upper limit of $\BR(\Bdmumu) < 3.4\times 10^{-10}$ at 95\%
confidence level (CL) which reduces the tension in this parameter.

The updated ATLAS measurement~\cite{ATLASBmumu2019} 
of the \Bsdmumu\ branching fractions
includes 36.2~\ifb\ of data taken at a centre-of-mass energy of 13~TeV
during 2015 and 2016 (LHC Run~2) and a combination with the
result based on 25~\ifb\ data
taken at 7-8~TeV during LHC Run~1.  For Run~2, events triggered by two muons
($p_\mathrm{T}(\mu_1) > 6$~GeV, $p_\mathrm{T}(\mu_2) > 4$~GeV, $|\eta| < 2.5$)
with the invariant di-muon mass \mmumu\ in the range of 4 to 8.5~GeV are
selected.
The dominant combinatorial background ($b\to\mu X \times
\bar{b}\to\mu X$ pairs) is rejected by a 15-variable Boosted Decision
Tree (BDT) which is trained and tested on data sidebands and simulated
signal events.
Tails from partially reconstructed $b\to \mumu X$ decays
like $B\to\mumu X$, $B\to c\mu X\to s(d)\mumu X$ or $B_c\to\Jpsi\mu\nu$,
which involve real di-muons at low \mmumu, and
semi-leptonic decays ($B_{(s)}/\Lambda^0_b\to h\mu\nu$ with
$h=\pi,K,p$) contribute to the signal region and are 
taken into account in the signal fit.
A small contribution of $B\to hh'$ ($h^{(\prime)}=\pi^\pm, K^\pm$) decays, 
with hadrons misidentified as muons,
peaks in the \Bsdmumu\ signal region
contributing $2.9\pm 2.0$ events after a ``tight'' muon
selection is applied.
The yield in the normalisation channel \BpmKpmJpsi\ with \JpsiMuMu\ is
determined by an unbinned maximum likelihood fit to \mJpsiK\ while the
efficiency relative to \Bsdmumu\ is extracted from Monte Carlo (MC) within
a fiducial volume defined by $p_T(B) > 8$~GeV and $|\eta_B| < 2.5$.
The overall efficiency ratio
$R_\varepsilon = \varepsilon_\JpsiK/\varepsilon_\mumu$ is
$0.1176\pm0.0009~\mathrm{(stat.)}\pm 0.0047~\mathrm{(syst.)}$
with the largest contribution to the systematic uncertainties
originating from data-MC discrepancies in the BDT input quantities.
A correction of 2.7\% has been applied to $R_\varepsilon$ to account
for the effective \Bs\ lifetime.

Due to the limited mass resolution the overlapping \Bs\ and
\Bd\ peaks are statistically separated by an unbinned maximum
likelihood fit to the \mmumu\ distributions in four BDT bins.  The
signal and \Bhh\ distributions are modelled by three double-Gaussian
PDFs, each with a common mean, while the background is described by a
first-order polynomial (combinatorial background) in combination with
an exponential distribution ($b\to\mumu X$ and semi-leptonic
background) whose shape parameters and normalisations are obtained
from data (Fig.~\ref{fig:bmumurun2mass4}).

\begin{figure}[t]
  \centerline{
    \includegraphics[width=80mm]{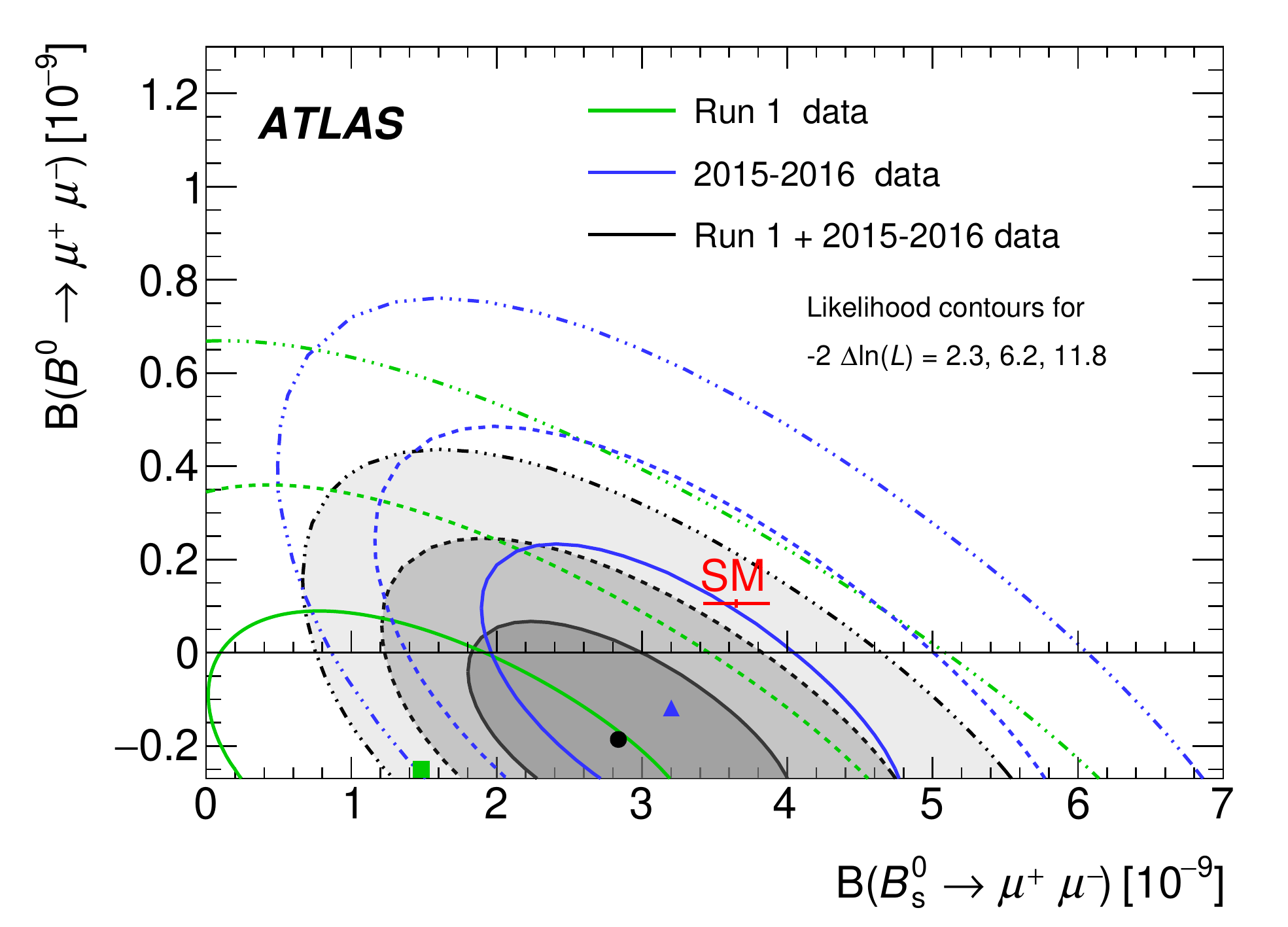}
  }
  \caption{\label{fig:bmumucombrun1run2}
    Likelihood contours for the combination of the Run~1 and 2015-2016
    Run~2 results (shaded areas). The contours are obtained with the
    combination of the two analyses likelihood, for values of
    $-2\,\DeltaLnLike$
    equal to 2.3, 6.2 and 11.8. The contours for the individual Run~2
    2015-2016 and Run~1 results are overlaid. The SM predictions and
    their uncertainties are
    included~\cite{ATLASBmumu2019}.
  }
\end{figure}

For the Run~2 data, yields of $N_s = 80\pm22$ \Bsmumu\ and
$N_d = -12\pm 20$ \Bdmumu\ events are extracted, consistent with SM
expectations of $N_s^{SM}=91$ and $N_d^{SM} = 10$, respectively.
Employing a Neyman construction (Fig.~\ref{fig:bmumurun2neym})
a branching fraction of $\BR(\Bsmumu) =
\left(3.21^{+0.96}_{-0.91}\;\mathrm{(stat.)}^{+0.49}_{-0.30}\;
\mathrm{(syst.)}\right) \times 10^{-9}$ and
an upper limit of
$\BR(\Bdmumu) < 4.3 \times 10^{-10}$ at 95\%~CL are
obtained.
A combination of the likelihood contours of the Run~2 (2015 and 2016)
and Run~1 results (Fig.~\ref{fig:bmumucombrun1run2}) is compatible with the
SM at $2.4~\sigma$ level and results in
$\BR(\Bsmumu) = \left(2.8^{+0.8}_{-0.7}\right) \times 10^{-9}$
and $\BR(\Bdmumu) < 2.1 \times 10^{-10}$ at 95\%~CL.

\section{CP-violation in \BsJpsiPhi}

\begin{figure}[t]
  \centerline{
    \includegraphics[width=\linewidth]{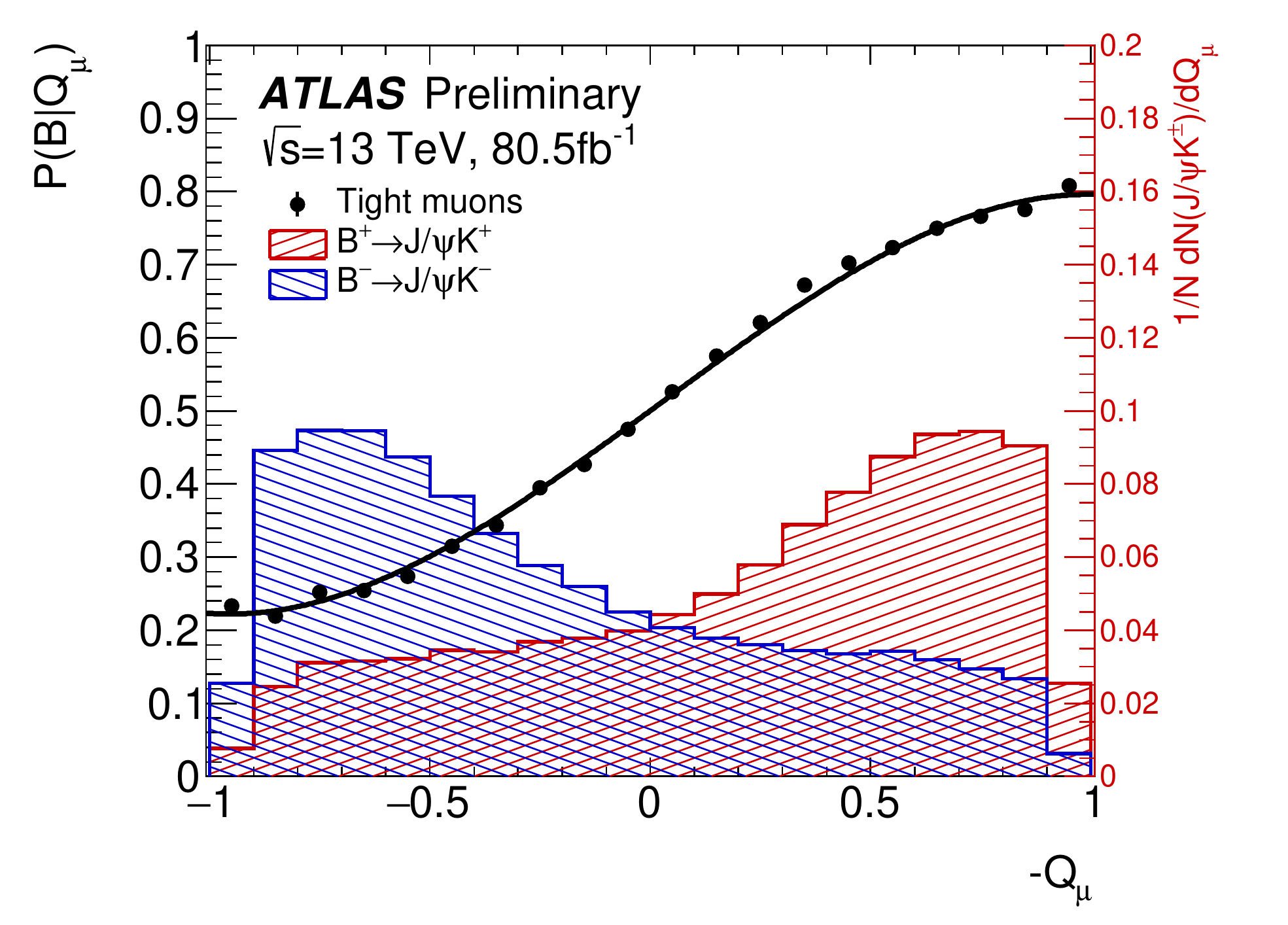}
  }
  \caption{\label{fig:bsjpsiphiconch2018}
   Cone charge distributions, $Q_\mu$, for tight muons, shown for the
   continuous distribution.  For each plot, in red (blue), the
   normalised $B^+$ ($B^−$) cone charge distribution is shown
   (corresponding to the right axis scale). Superimposed is the
   distribution of the tag-probability, $P(B|Q_\mu)$, as a function of
   the cone charge, derived from a data sample of \BpmKpmJpsi\, and
   defined as the probability to have a $B^+$ meson (on the
   signal-side) given a particular cone charge $Q_\mu$.
   The fitted parametrization, shown in black, is used as the
   calibration curve to infer the probability to have a \Bs\ or
   \Bsbar\ meson produced at production in the decays
   to \BsJpsiPhi~\cite{ATLASBsJpsiPhi2019}.
  }
\end{figure}

In the SM, the CP violating phase $\phi_s$ in the \BsJpsiPhi\ decay
(with \JpsiMuMu\ and \Phikk) is small and can be predicted to
$\phi_s \approx -2\,\beta_s = -0.0363^{+0.016}_{-0.0015}$~rad~\cite{Charles2011}.
The ATLAS Run~1 measurement~\cite{atlasbsjpsiphi2016} of
$\phi_s = -0.090\pm 0.078~\mathrm{(stat.)}\pm
0.041~\mathrm{(syst.)}$~rad and of the decay width difference
$\Delta\Gamma_s = 0.085 \pm 0.011~\mathrm{(stat.)}\pm
0.007~\mathrm{(syst.)}~\mathrm{ps}^{-1}$ agrees with SM expectations
($\Delta\Gamma_s = 0.087\pm 0.021~\mathrm{ps}^{-1}$ in
SM~\cite{LenzNierste2011}) and
is consistent with results from other experiments.

The ATLAS Run-2 \BsJpsiPhi\ measurement~\cite{ATLASBsJpsiPhi2019}
uses 80.5~\ifb\ of 13~TeV data
taken in 2015-2017 selected by multiple triggers based on
\JpsiMuMu decays with muon-\pt\ thresholds of 4 or 6~GeV.
In order to extract the flavour of the decaying \Bs\ (or \Bsbar)
opposite-side taggers which rely on the \pt-weighted charge of tracks
$Q_x$ 
inside a cone around either an electron, a muon or a $b$-jet are used.
The taggers are calibrated on self-tagging \BpmKpmJpsi\ events and
yield a total tagging power of $1.65\pm 0.01$\% with the tight muon
tagger contributing $0.86\pm 0.01$\%, about half of the tagging
power (Fig.~\ref{fig:bsjpsiphiconch2018}).

\begin{figure}[t]
  \centerline{
    \includegraphics[width=\linewidth]{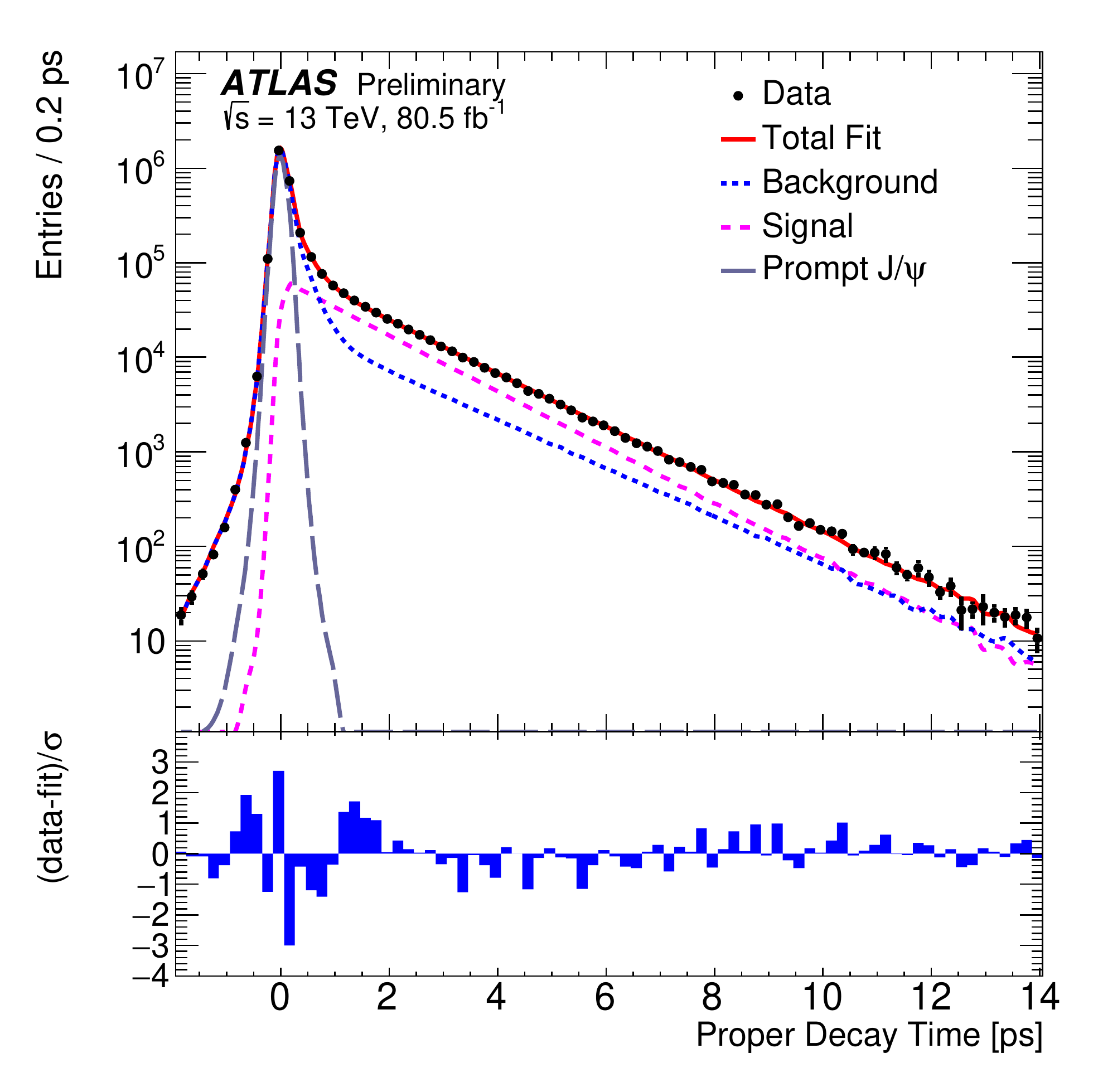}
  }
  \caption{\label{fig:bsjpsiphiproptfit2018}
    Proper decay time fit projection for the \BsJpsiPhi\ sample. The red
    line shows the total fit, while the magenta dashed line shows the
    total signal. The total background is shown as a blue dashed line
    with a long-dashed grey line showing the prompt \Jpsi\
    background. Below the figure is a ratio plot that shows the
    difference between each data point and the total fit line divided
    by the statistical and systematic uncertainties summed in
    quadrature of that point~\cite{ATLASBsJpsiPhi2019}.
  }
\end{figure}

\begin{figure}[t]
  \centerline{
    \includegraphics[width=\linewidth]{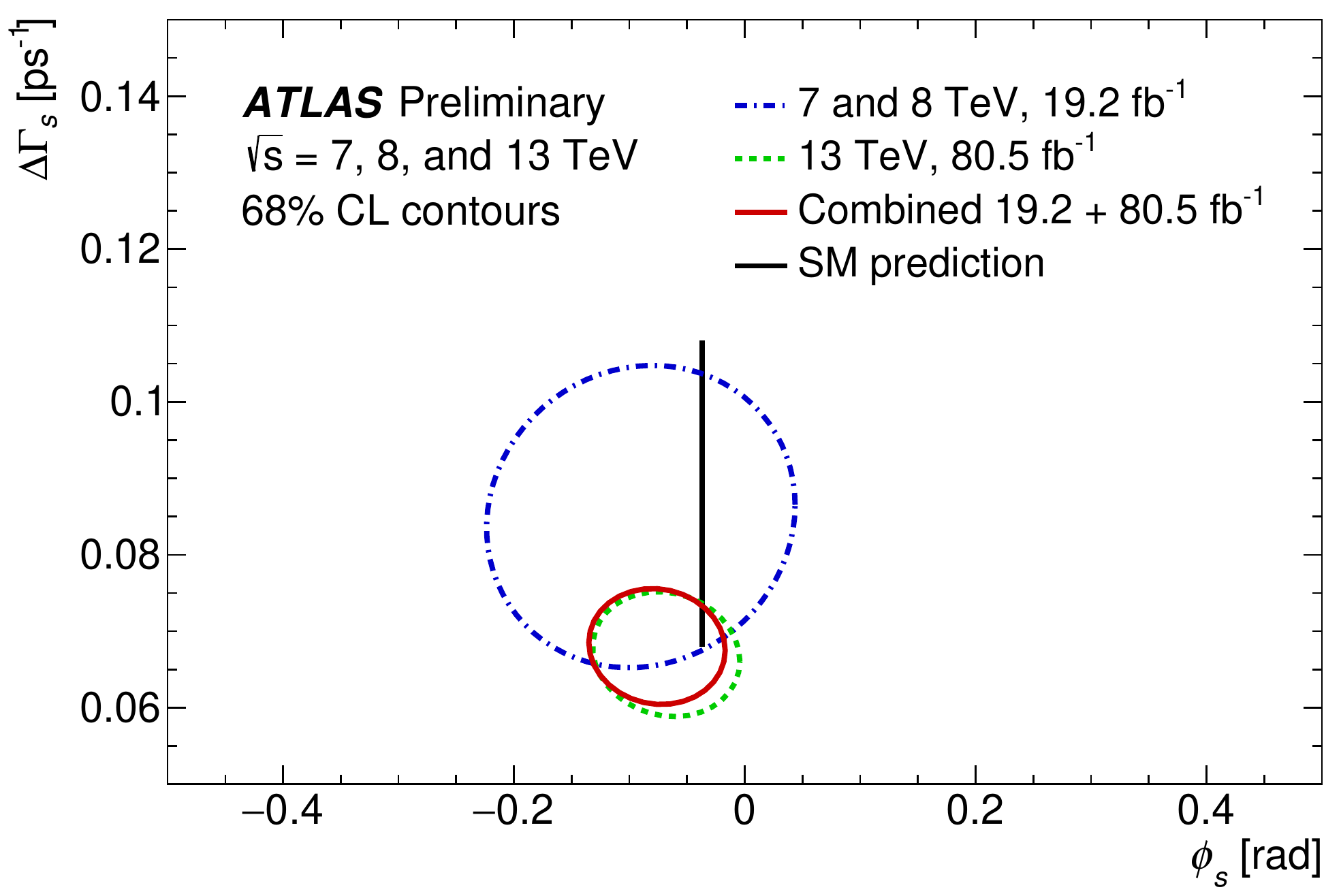}
  }
  \caption{\label{fig:bsjpsiphires2018}
    Likelihood 68\% confidence level contours in the
    $\phi_s$ - $\Delta\Gamma_s$ plane, showing ATLAS results for 7~TeV
    and 8~TeV data (blue dashed-dotted curve), for 13~TeV data
    (green dashed curve) and for 13~TeV data combined with 7~TeV and
    8~TeV (red solid curve) data. In all contours the statistical and
    systematic uncertainties are combined in quadrature and
    correlations are taken into
    account~\cite{ATLASBsJpsiPhi2019}.
  }
\end{figure}

An unbinned maximum likelihood fit based on the \Bs\ properties
(\pt\, mass
and mass uncertainty, proper decay time
(Fig.~\ref{fig:bsjpsiphiproptfit2018})
and uncertainty,
the \Bs-flavour tagging probability $p(B|Q_x)$) and the transversity
angles $\Omega(\theta_T, \phi_T, \psi_T)$, defined in
\cite{ATLASBsJpsiPhi2019}, is employed to extract nine signal parameters.
For Run~2 data only, values of 
$\phi_s = -0.068\pm 0.038~\mathrm{(stat.)}\pm 0.018~\mathrm{(syst.)}$~rad
and
$\Delta\Gamma_s  = -0.067\pm 0.005~\mathrm{(stat.)}\pm
0.002~\mathrm{(syst.)~ps^{-1}}$
(Fig.~\ref{fig:bsjpsiphires2018})
are obtained.

\begin{figure}[t]
  \centerline{
    \includegraphics[width=\linewidth]{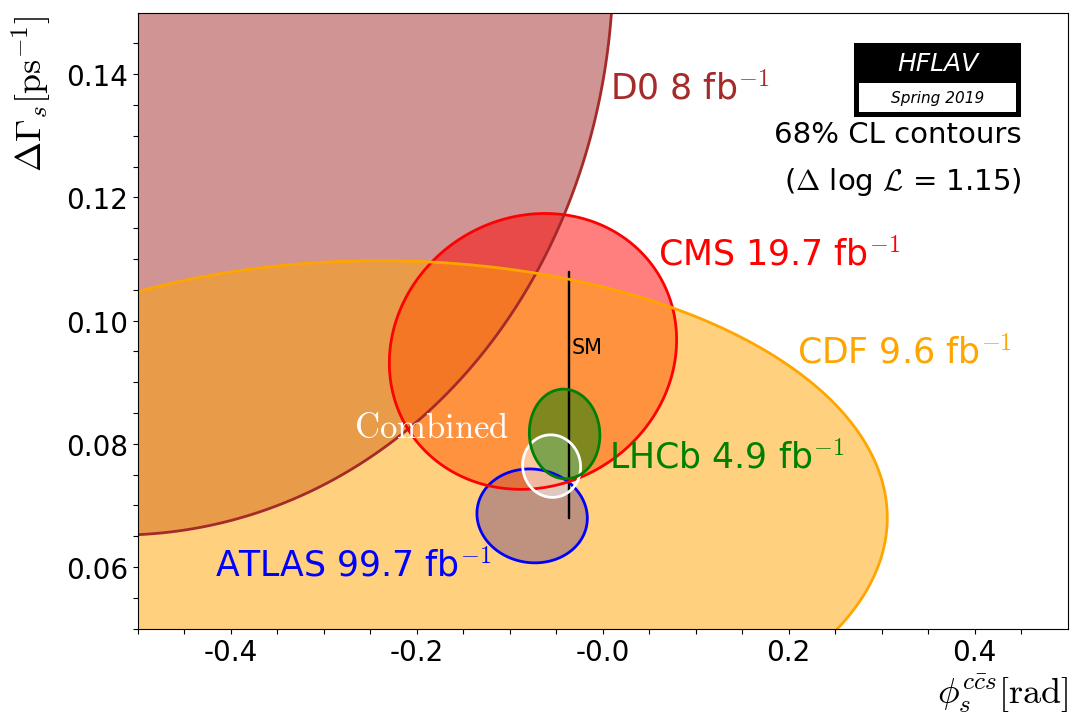}
  }
  \caption{\label{fig:bsjpsiphihflav201905}
    Likelihood 68\% confidence level contours in the $\phi_s$ -
    $\Delta\Gamma_s$  plane, including results from D0 (brown),
    CDF (yellow), LHCb (green) and CMS (red). The brown contour with blue
    edge shows the ATLAS result for 13~TeV combined with 7~TeV and
    8~TeV.  The statistical and systematic uncertainties are combined
    in quadrature.  A preliminary HFLAV combination is shown by the
    white contour~\cite{hflavbsjpsiphispring2019}.
  }
\end{figure}

The combined ATLAS Run~1 and Run~2 result yields
$\phi_s = -0.076\pm 0.034~\mathrm{(stat.)}\pm 0.019~\mathrm{(syst.)}$~rad
and
$\Delta\Gamma_s  = -0.068\pm 0.004~\mathrm{(stat.)}\pm
0.003~\mathrm{(syst.)~ps^{-1}}$
which are consistent with the SM expectations as well as results from
other experiments
(Fig.~\ref{fig:bsjpsiphihflav201905}).
A preliminary HFLAV average~\cite{hflavbsjpsiphispring2019} results in
$\phi_s = -0.055\pm 0.021$~rad
and
$\Delta\Gamma_s  = -0.0764^{+0.0034}_{-0.0033}~\mathrm{ps}^{-1}$.

\newpage

\onecolumngrid

\begin{figure}[t]
  \centerline{
    \begin{overpic}[width=0.5\textwidth]{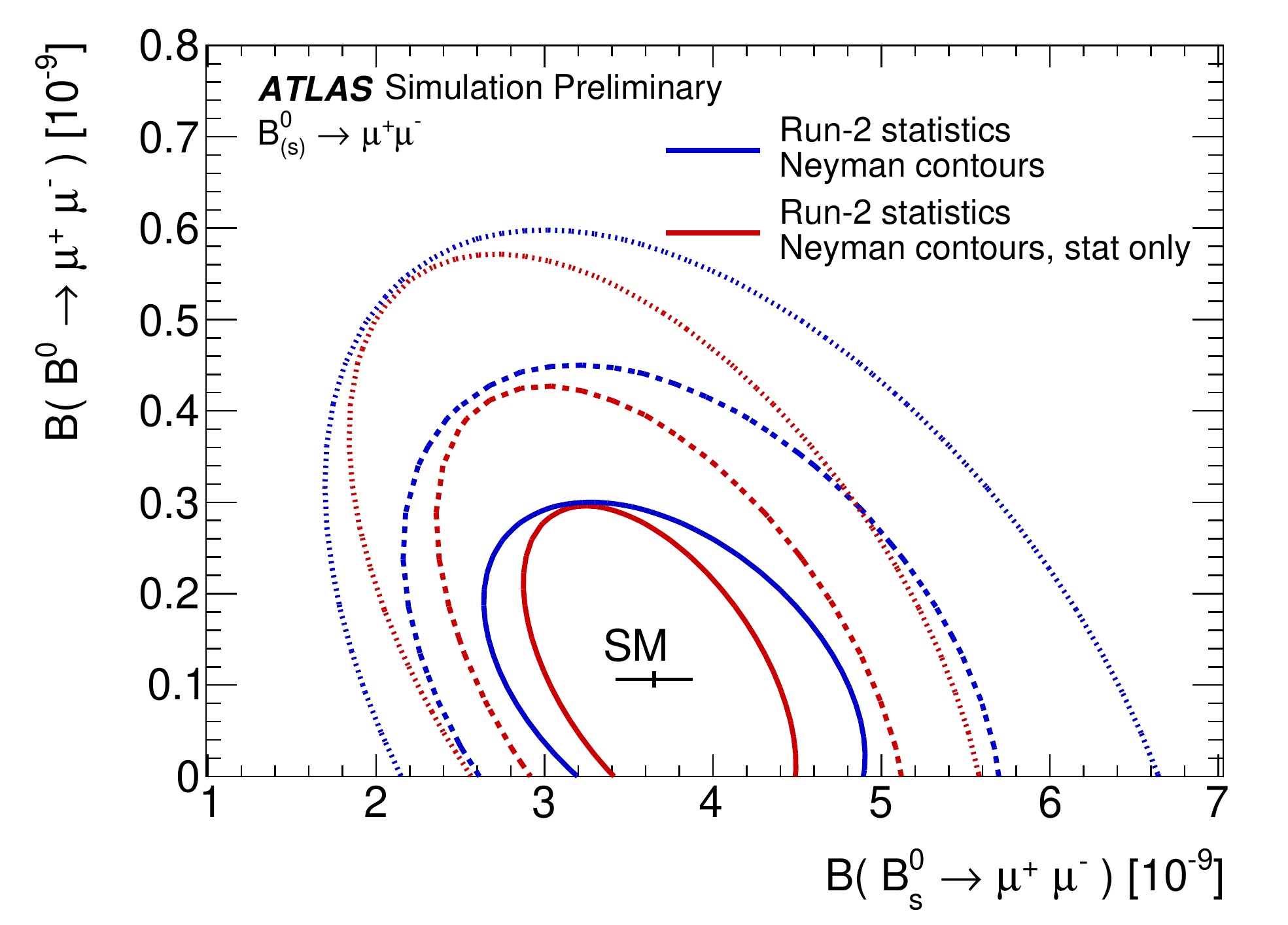}%
      \put(20,15){(a)}
    \end{overpic}
    \hfill%
    \begin{overpic}[width=0.5\textwidth]{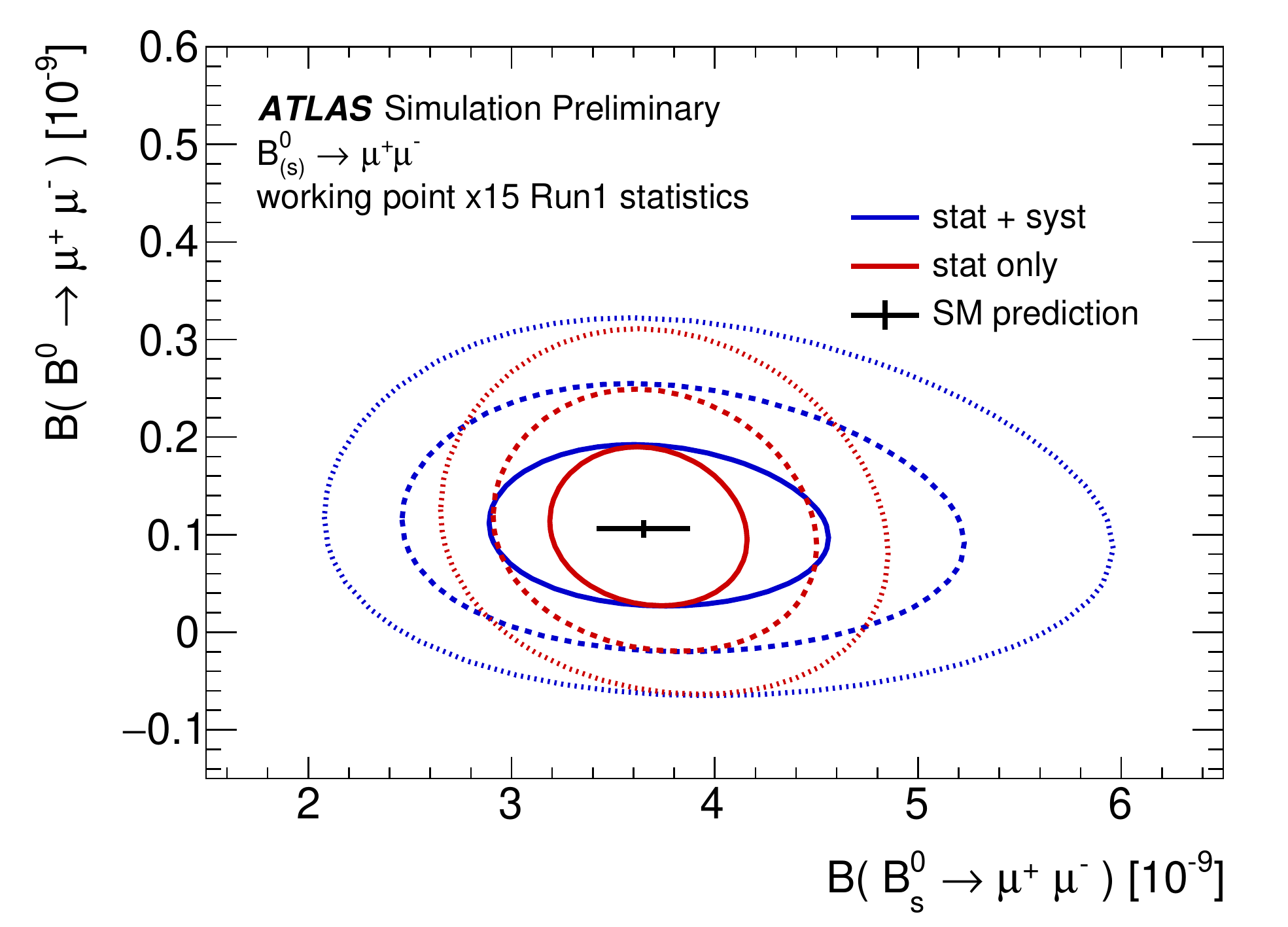}%
      \put(20,15){(b)}
    \end{overpic}
  }
  \centerline{
    \begin{overpic}[width=0.5\textwidth]{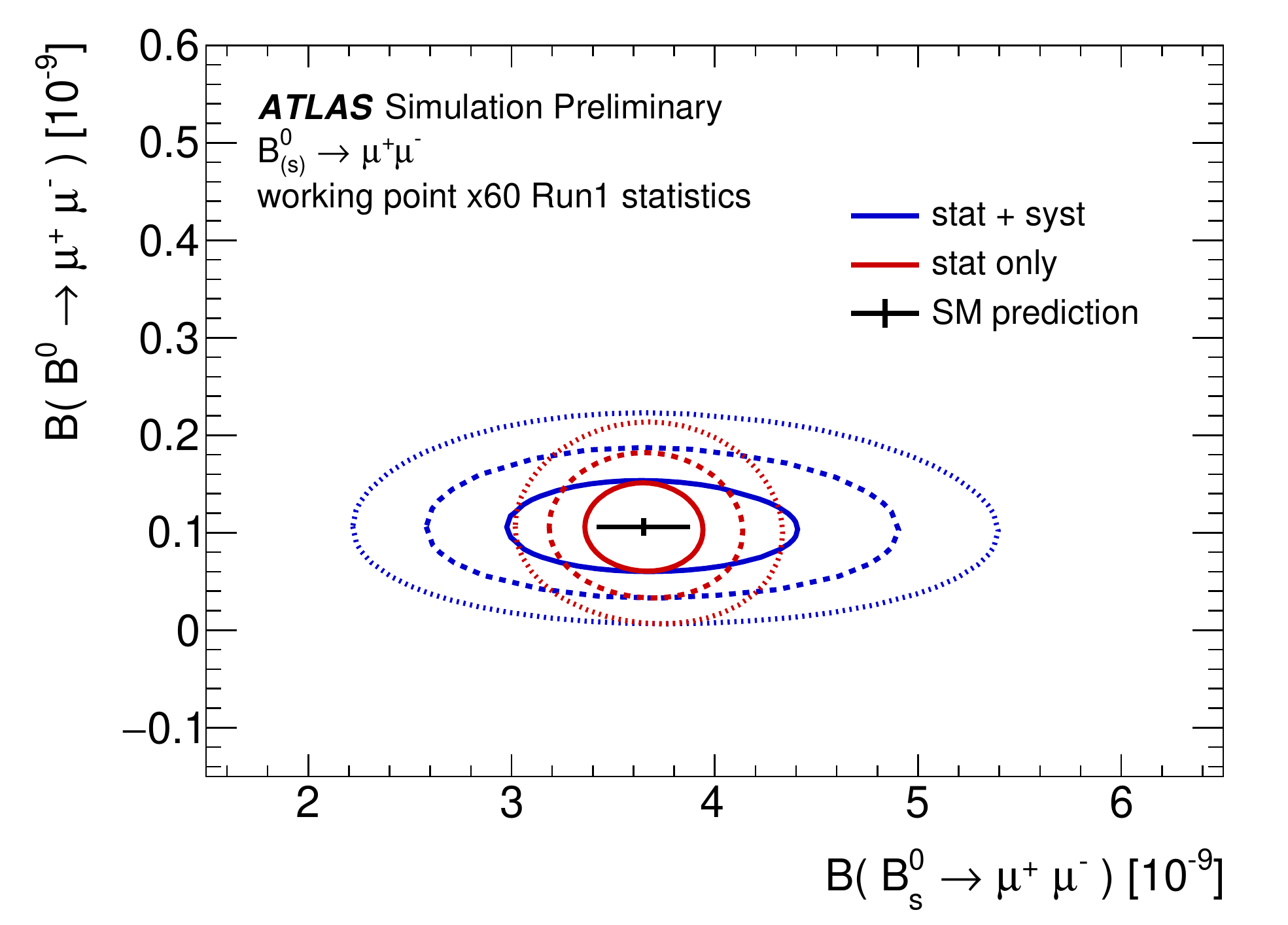}%
      \put(20,15){(c)}
    \end{overpic}
    \hfill%
    \begin{overpic}[width=0.5\textwidth]{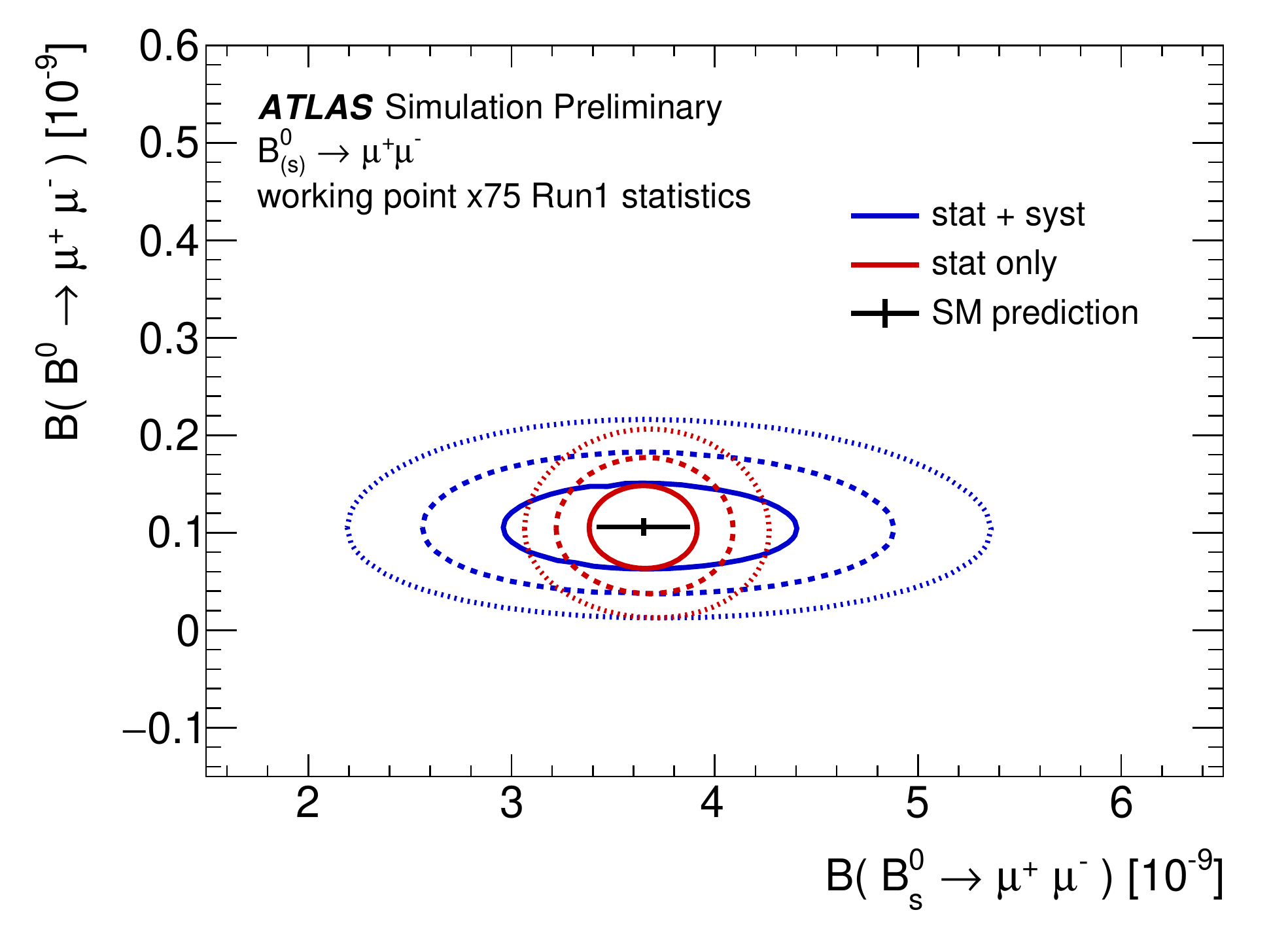}%
      \put(20,15){(d)}
    \end{overpic}
  }
  \caption[]{\label{fig:bmumuprojections}
    {(a):}
    Comparison of 68.3\% (solid), 95.5\% (dashed) and 99.7\% (dotted)
    confidence level contours obtained exploiting the 2D Neyman belt
    construction for the full LHC Run~2 case~\cite{atlasbmumuproj2018}.
    Red contours are statistical only; blue contours
    include systematics uncertainties from the ATLAS Run~1
    analysis~\cite{atlasbs2016} extrapolated to Run~2 statistics.
    The black points show the SM
    theoretical prediction and its uncertainty~\cite{bobeth2014}.\newline
    {(b) - (d):}
    Comparison of confidence level profiled likelihood ratio
    contours for {(b)} the ``conservative'',
    {(c)} the ``intermediate'' and
    {(d)} the ``high-yield'' HL-LHC extrapolation with
    $\times 15$, $ \times 60$ and $\times 75$ the Run~1 statistics
    for the (10~GeV, 10~GeV), the (6~GeV, 10~GeV) and the
    (6~GeV, 6~GeV) dimuon trigger scenarios,
    respectively~\cite{atlasbmumuproj2018}.
  }
\end{figure}
\twocolumngrid


\section{High-luminosity LHC prospects}

The branching fraction measurement of the very rare decays
\Bsmumu\ and \Bdmumu
will benefit from the
increased statistics and the improved invariant mass resolution at the
HL-LHC.
The separation of the \Bs\ and \Bd\ mass peaks increases by a factor of 
1.65 (1.5) to $2.3\,\sigma$ ($1.3\,\sigma$) in the
barrel (end-cap) region compared to Run~1~\cite{atlasHLperformance2016}.

The projection of the ATLAS detector performance for measuring
\BR(\Bsdmumu) with the expected datasets during the full LHC
Run~2 (130~\ifb) and at the HL-LHC
(3\,000~\ifb)~\cite{atlasbmumuproj2018} 
using pseudo-MC experiments is based on the likelihood of the Run~1
analysis.  The signal statistics estimate for the Run~2 scenario
applies scaling factors for the integrated luminosity, the
cross-section increase due to the higher center-of-mass energy of
13~TeV and the muon pair selection with topological triggers
with ($\pt(\mu_{1,2}) > 6$~GeV) or ($\pt(\mu_1) > 6$~GeV, $\pt(\mu_2) >
4$~GeV) thresholds resulting in 7 times the number of signal events in
Run~1.  The contours of the 2-dimensional Neyman construction
(Fig.~\ref{fig:bmumuprojections} {\em (a)}) include the external
systematic uncertainties on the $b$-quark fragmentation fractions 
$f_s/f_d$ and \BR(\BpmKpmJpsi)  which were
kept the same as in the Run~1 analysis as well as internal ones like
the fit shapes and efficiencies which were scaled according to the
increase in statistics.
For the HL-LHC case three potential trigger scenarios are considered:
two muons with $\pt > 10$~GeV (``conservative''),
one muon with $\pt > 10$~GeV and another with $\pt > 6$~GeV
(``intermediate'') 
as well as two muons with $\pt > 6$~GeV (``high yield'')  
providing 15, 60 and 75 times the Run~1 statistics, respectively.
The profile likelihood contours of pseudo-experiments based again on
the likelihood of the Run~1 analysis demonstrate the increased sensitivity 
of the ATLAS detector for \BR(\Bsmumu) and \BR(\Bdmumu) at the HL-LHC
(Fig.~\ref{fig:bmumuprojections} {\em (b)-(d)}).  The uncertainty on
the $f_s/f_d$ value, conservatively taken as 8.3\%
from the ATLAS measurement~\cite{atlasfsfd2015},
dominates the systematic uncertainty contributions on \BR(\Bsmumu).

\begin{figure}[t]
  \centerline{
    \includegraphics[width=\linewidth]{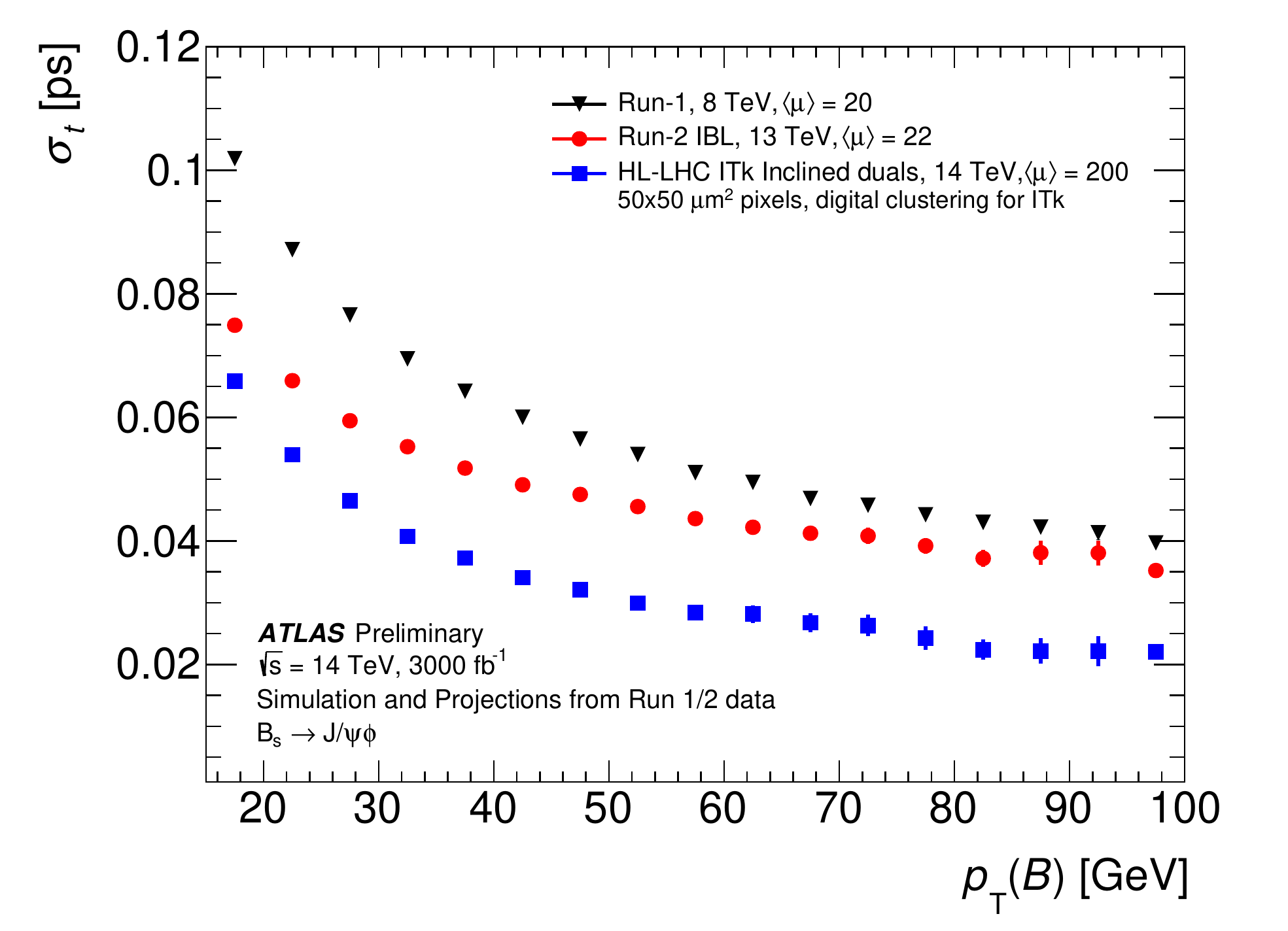}
  }
  \caption{\label{fig:bsjpsiphiHLpropt2019}
    Dependence of the proper decay time resolution of the \Bs\ meson
    of the signal \BsJpsiPhi\ decay on \Bs\ \pt.
    Per-candidate resolutions corrected for scale-factors are shown,
    comparing the performance in Run~1 (ID), Run~2 (IBL) and
    upgrade HL-LHC (ITk) MC simulations. All samples use 6~GeV muon \pt\
    cuts~\cite{atlasbsjpsiphipros2018}.
  }
\end{figure}

The prospects of measuring the CP-violating phase $\phi_s$ and the
\Bs\ decay width difference $\Delta\Gamma_s$ using \BsJpsiPhi\ decays
at the HL-LHC ($3\,000~\ifb$)~\cite{atlasbsjpsiphipros2018} have
been explored by pseudo-MC experiments based on the Run~1
\BsJpsiPhi\ analysis using similar trigger scenarios as in the
\Bsdmumu\ HL-LHC study, yielding a \Bs\ signal statistics increase
of $\times 18$, $\times 60$ and $\times 100$ w.r.t. the yield obtained
in 2012 data for the ``conservative'', ``intermediate'' and
``high-yield'' scenarios, respectively.
The sensitivity to $\phi_s$ as well as to $\Delta\Gamma_s$ is improved
considerably by the detector upgrades, especially the proper time
resolution $\sigma_t$ (Fig.~\ref{fig:bsjpsiphiHLpropt2019}).
In the calculation of the expected uncertainties on $\phi_s$ and
$\Delta\Gamma_s$ the number of \Bs\ signal events and the
proper time resolution $\sigma_t$ are assumed to scale with the
integrated luminosity while the \Bs\ flavour tagging power
-- conservatively -- is not scaled.
The systematic uncertainties (likelihood fit model description,
\Bs\ flavor tagging calibration, detector acceptance description,
detector alignment, peaking background contributions) are expected to
improve with increased statistics as well, providing estimates of
$\delta_{\phi_s}^{syst} \approx 0.003$~rad
and $\delta_{\Delta\Gamma_s}^{syst} \approx 0.0005~\mathrm{ps}^{-1}$
for an integrated luminosity of $3\,000~\ifb$.
The improvement in the statistical uncertainties obtained w.r.t.
the Run~1 result are factors 9 to 20 for $\phi_s$, 
up to 7 times smaller than the SM prediction for $\phi_s$, 
and factors 4 to 10 for $\Delta\Gamma_s$.
The 68\% CL contours for the three scenarios  
(Fig.~\ref{fig:bsjpsiphiHLcont2019}) include the combination
of statistical and systematic uncertainties.

\begin{figure}[t]
  \centerline{
    \includegraphics[width=\linewidth]{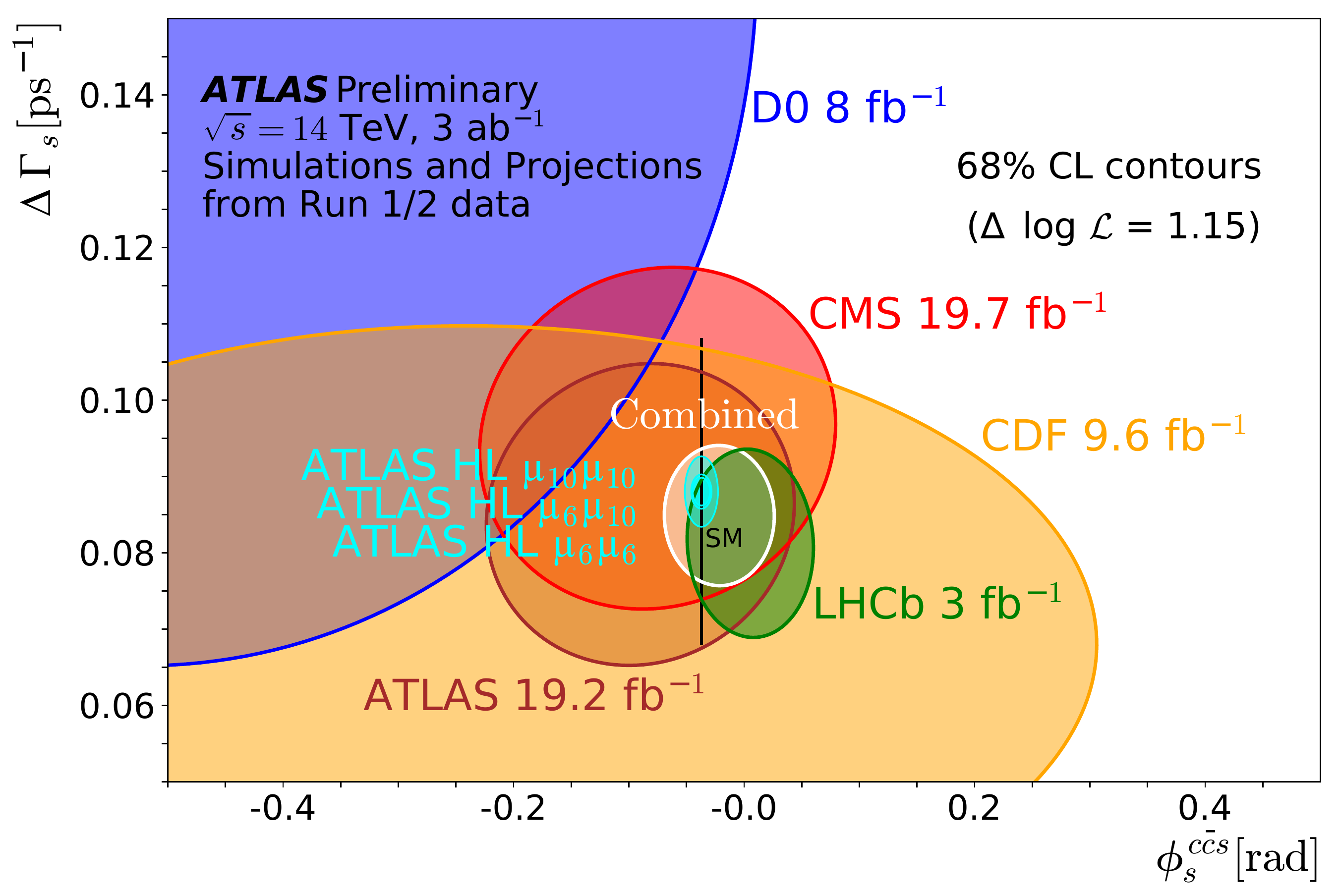}
  }
  \caption{\label{fig:bsjpsiphiHLcont2019}
    Experimental summary of the $\phi_s$ - $\Delta\Gamma_s$
    measurements with superimposed ATLAS HL-LHC extrapolations,
    including both the projected statistical and systematic
    uncertainties~\cite{atlasbsjpsiphipros2018}.
  }
\end{figure}

\section{Summary}

Measurements of rare decays and CP-violation by the ATLAS
collaboration have been presented.
The results for
$\BR(\Bsmumu)$ and the search for the decay
$\BR(\Bdmumu)$ with 36.2~\ifb\ of Run-2 data agree with the
Standard Model and other measurements.  There is no
sign for the decay \Bdmumu\ in ATLAS data,
but ATLAS will add approximately data taken in 2017 and 2018
to the analysis ($\approx 107~\ifb$).

The ATLAS measurement of the CP-violating phase $\phi_s$ and
the \Bs\ decay width difference $\Delta\Gamma_s$ provides a single
measurement precision comparable to that of the LHCb experiment and
reaches the sensitivity to test the Standard Model prediction.
About 60~\ifb\ of data taken in 2018 will be added to the analysis
in the future.

Both analyses will profit considerably from the increased
statistics expected from the $3\,000~\ifb$ of HL-LHC data as well as
detector improvements providing better mass and proper decay time
resolutions. This will allow more stringent tests of the Standard Model.

\begin{acknowledgments}
  This work was partially supported by grants of the German Federal
  Ministry  of Education and Research (BMBF) and the German Helmholtz
  Alliance ``Physics at the Terascale''.
\end{acknowledgments}

\bigskip 


\begin{thebibliography}{99} 

\bibitem{ATLASdet} ATLAS Collaboration, 
  \emph{The ATLAS Experiment at the CERN Large Hadron Collider}, 
  2008 JINST 3 S08003
\bibitem{LHCpaper} L. Evans and P. Bryant (editors),
  \emph{LHC Machine},
  2008 JINST 3 S08001
\bibitem{ybHLLHC}
  G. Apollinari, I. B\'ejar Alonso, O. Br\"uning, P. Fessia,
  M. Lamont, L. Rossi, L. Tavian (editors),
  {\em High-Luminosity Large Hadron Collider (HL-LHC),
    Technical Design Report V. 0.1}, 
  CERN Yellow Reports Vol. 4/2017,
  CERN-2017-007-M (CERN, Geneva, 2017)
  [\href{https://cds.cern.ch/record/2284929}
    {{\texttt{https://cds.cern.ch/record/2284929}}}]
\bibitem{bobeth2014} C. Bobeth et al.,
  {\em $\Bsd \to \ell^+\ell^-$ in the Standard Model with Reduced
    Theoretical Uncertainty},
  Phys. Rev. Lett. 112 (2014) 101801
\bibitem{atlasbs2016} ATLAS Collaboration, 
  {\em Study of the rare decays of \Bs\ and \Bd\ into muon pairs from
    data collected during the LHC Run~1 with the ATLAS detector},
  Eur.~Phys.~J.~C (2016) 76:513
\bibitem{CMSLHCbcomb2015} CMS and LHCb Collaborations,
  {\em Observation of the rare \Bsmumu\ decay from the combined 
    analysis of CMS and LHCb data}, 
  Nature, {\bf 522}, 2015
\bibitem{LHCbBmumu2017} LHCb Collaboration,
  {\em Measurement of the \Bsmumu\ Branching Fraction and Effective
    Lifetime and Search for \Bdmumu\ Decays},
  Phys.~Rev.~Lett. 118 (2017) 191801
\bibitem{ATLASBmumu2019} ATLAS Collaboration,
  \emph{Study of the rare decays of \Bs\ and \Bd\ mesons into
    muon pairs using data collected during 2015 and 2016 with the
    ATLAS detector},
  JHEP04 (2019) 098
\bibitem{Charles2011} J. Charles et al.,
  \emph{Predictions of selected flavour observables within the
    standard model},
  Phys. Rev. D {\bf 84} (2011) 033005
\bibitem{LenzNierste2011} A. Lenz and U. Nierste,
  \emph{Numerical updates of lifetimes and mixing parameters of B
    mesons},
  (2011), {{\texttt{arXiv:1102.4274}}}
\bibitem{atlasbsjpsiphi2016} ATLAS Collaboration,
  {\em Measurement of the CP-violating phase $\phi_s$ and the $\Bs$ meson
    decay width difference with $\BsJpsiPhi$ decays in ATLAS},
  JHEP~08~(2016)~147
\bibitem{ATLASBsJpsiPhi2019} ATLAS Collaboration,
  \emph{Measurement of the CP violation phase $\phi_s$ in
    \BsJpsiPhi\ decays in ATLAS at 13~TeV},
  ATLAS-CONF-2019-009,
  [\href{https://cds.cern.ch/record/2668482}
    {{\texttt{https://cds.cern.ch/record/2668482}}}]
\bibitem{hflavbsjpsiphispring2019} Heavy Flavor Averaging Group,
  {Preliminary combination of $\phi_s$ vs. $\Delta\Gamma_s$ results,
    spring 2019},
  CERN seminar presentation by F. Dordei, 2019-05-07,
  [\href{https://hflav.web.cern.ch/}
    {{\texttt{https://hflav.web.cern.ch/}}}]  
\bibitem{atlasHLperformance2016} ATLAS Collaboration,
  {\em Expected peformance for an upgraded ATLAS detector at
    High-Luminosity LHC},
  ATL-PHYS-PUB-2016-026
  [\href{https://cds.cern.ch/record/2223839}
    {{\texttt{https://cds.cern.ch/record/2223839}}}]
\bibitem{atlasbmumuproj2018} ATLAS Collaboration,
  {\em Prospects for the \BR(\Bsdmumu) measurements 
    with the ATLAS detector in Run~2 LHC and HL-LHC data campaigns},
  ATL-PHYS-PUB-2018-005, 
  [\href{https://cds.cern.ch/record/2317211}
    {{\texttt{https://cds.cern.ch/record/2317211}}}]
\bibitem{atlasfsfd2015} ATLAS Collaboration,
  {\em Determination of the ratio of $b$-quark fragmentation
    fractions $f_s/f_d$ in $pp$ collisions at $\sqrt{s} = 7$~TeV with the
    ATLAS detector},
  Phys. Rev. Lett. 115, 262001 (2015)
\bibitem{atlasbsjpsiphipros2018} ATLAS Collaboration,
  \emph{CP-violation measurement prospects in the \BsJpsiPhi\ channel
    with the upgraded ATLAS detector at the HL-LHC},
  ATLAS-PHYS-PUB-2018-041, 
  [\href{https://cds.cern.ch/record/2649881}
    {{\texttt{https://cds.cern.ch/record/2649881}}}]
  
\end{thebibliography}


\end{document}